\documentclass[12pt]{iopart}
\usepackage{graphics}
\usepackage{graphicx}
\usepackage[sort&compress,numbers]{natbib}

\begin{document}

\title[Electronic Structure of a Recently Synthesized Graphene-like BCN Monolayer]{On the Electronic Structure of a Recently Synthesized Graphene-like BCN Monolayer from bis-BN Cyclohexane: A DFT Study}

\author{Ramiro M. dos Santos$^1$, William F. Giozza$^2$, Rafael T. de Sousa J\'unior$^2$, Dem\'etrio A. da Silva Filho$^1$, and Luiz A. Ribeiro J\'unior$^{1,*}$}

\address{$^{1}$ \quad Institute of Physics, University of Bras\'ilia, Bras\'ilia, 70910-900, Brazil}
\address{$^{2}$ \quad Department of Electrical Engineering, University of Bras\'{i}lia 70919-970, Brazil}
\ead{ribeirojr@unb.br}
\vspace{10pt}
\begin{indented}
\item[]December 2020
\end{indented}

\begin{abstract}
Since the rising of graphene, boron nitride monolayers have been deeply studied due to their structural similarity with the former. A hexagonal graphene-like boron-carbon-nitrogen (h-BCN) monolayer was synthesized recently using bis-BN cyclohexane (B$_2$N$_2$C$_2$H$_{12}$) as a precursor molecule. Herein, we investigated the electronic and structural properties of this novel BCN material, in the presence of single-atom (boron, carbon, or nitrogen) vacancies, by employing density functional theory calculations. The stability of these vacancy-endowed structures is verified from cohesion energy calculations. Results showed that a carbon atom vacancy strongly distorts the lattice leading to breaking on its planarity and bond reconstructions. The single-atom vacancies induce the appearance of flat midgap states. A significant degree of charge localization takes place in the vicinity of these defects. It was observed a spontaneous magnetization only for the boron-vacancy case, with a magnetic dipole moment about 0.87 $\mu_B$. Our calculations predicted a direct electronic bandgap value of about 1.14 eV, which is in good agreement with the experimental one. Importantly, this bandgap value is intermediate between gapless graphene and insulating h-BN.
\end{abstract}

\section{Introduction}
Layered 2D materials have been promoted significant advances in the field of flat optoelectronics aiming at designing devices with good cost-efficiency compromise \cite{sun2016optical,gupta2015recent,zhang2016van}. These materials have gained much attention since the first synthesis of graphene \cite{novoselov2004electric,geim2009graphene,geim2007rise}. Graphene and its allotropes \cite{enyashin2011graphene,hirsch2010era} are still the focus of investigation mostly due to their performance in the operation of organic-based energy conversion and storage applications \cite{pumera2011graphene,bonaccorso2015graphene}. The combination of their unique mechanical and electronic properties allowed the production of large-area films, which are desirable for several technological applications \cite{li2009large,eda2008large}. Materials with a similar structure to the graphene, such as hexagonal boron nitride (h-BN) \cite{song2010large} and MoS$_2$ \cite{splendiani2010emerging}, have also become protagonists in designing novel optoelectronic applications. These materials are attractive alternatives to graphene since they have non-null bandgaps \cite{song2010large,splendiani2010emerging}.           

Recently, two-dimensional boron–nitrogen–carbon monolayers (h-BCN) with tunable direct bandgaps have been both experimentally \cite{liu2013plane,tran2016quantum,wu2012nitrogen,zhao2013local,maeda2017orientation} and theoretically \cite{zhang2015two,azevedo2007theoretical,korona2019exploring,gomes2013stability} studied. Their structures consisted of an in-plane combination of homogeneous graphene and h-BN sheets or nanoribbons. The electronic and structural properties of in-plane graphene/h-BN heterojunctions (2D h-BCN alloy monolayers) were systematically studied by adopting a computational protocol that combined a global optimization method (CALYPSO) and density functional theory (DFT) calculations \cite{zhang2015two}. The results revealed that h-BCN monolayers possess direct bandgaps ranging between 1.04-1.60 eV, depending on the resulting stoichiometries from the CALYPSO \cite{zhang2015two}. Through an experimental investigation, it was demonstrated the creation of 2D in-plane graphene/h-BN heterojunctions with controlled domain sizes by using lithography patterning and sequential CVD growth steps \cite{liu2013plane}. Through this technique, the shapes of graphene and h-BN domains were controlled precisely, and sharp graphene/h-BN interfaces were created. 

The search for novel h-BCN structures has motivated the development of new experimental techniques capable of yielding boron nitride-based materials with semiconducting bandgaps \cite{GONZALEZORTIZ2020100107,beniwal2017graphene}. Beniwal and coworkers proposed a strategy to synthesize a 2D graphetic but ternary monolayer containing atoms of carbon, nitrogen, and boron \cite{beniwal2017graphene}. The strategy utilizes bis-BN cyclohexane (B$_2$N$_2$C$_2$H$_{12}$) as a precursor molecule and relies on thermally induced dehydrogenation of them and the formation of an epitaxial monolayer on Ir(111) through covalent bond formation \cite{beniwal2017graphene}. The ternary monolayers obtained with this method have good thermal stability and semiconducting character. Importantly, their experimental findings showed remarkable similarities between the h-BCN and the previously reported h-BN and graphene monolayers deposited on various transition metal substrates \cite{beniwal2017graphene}. To pave the way for the development of future flexible electronic and optical devices based on these novel ternary h-BCN monolayers, the structural and electronic properties of their defective sheets should be deeply understood. To the best of our knowledge, such an investigation has not been performed so far.      

Herein, motivated by the recent synthesis of ternary h-BCN monolayers \cite{beniwal2017graphene}, DFT calculations were employed to study the electronic and structural properties of these materials in the presence of vacancy defects. Our computational protocol considers three distinct model defective lattices with a single-atom vacancy by removing a carbon, boron, or nitrogen atom. For comparison purposes, the results obtained for these systems are contrasted with the ones for the non-defective lattice. This approach allowed us to demonstrate how a single-atom vacancy plays a role in enabling magnetization effects in ternary h-BCN materials.

\section{Details of Modeling}

\begin{figure}[!htb]
    \centering
    \includegraphics[width=0.7\linewidth]{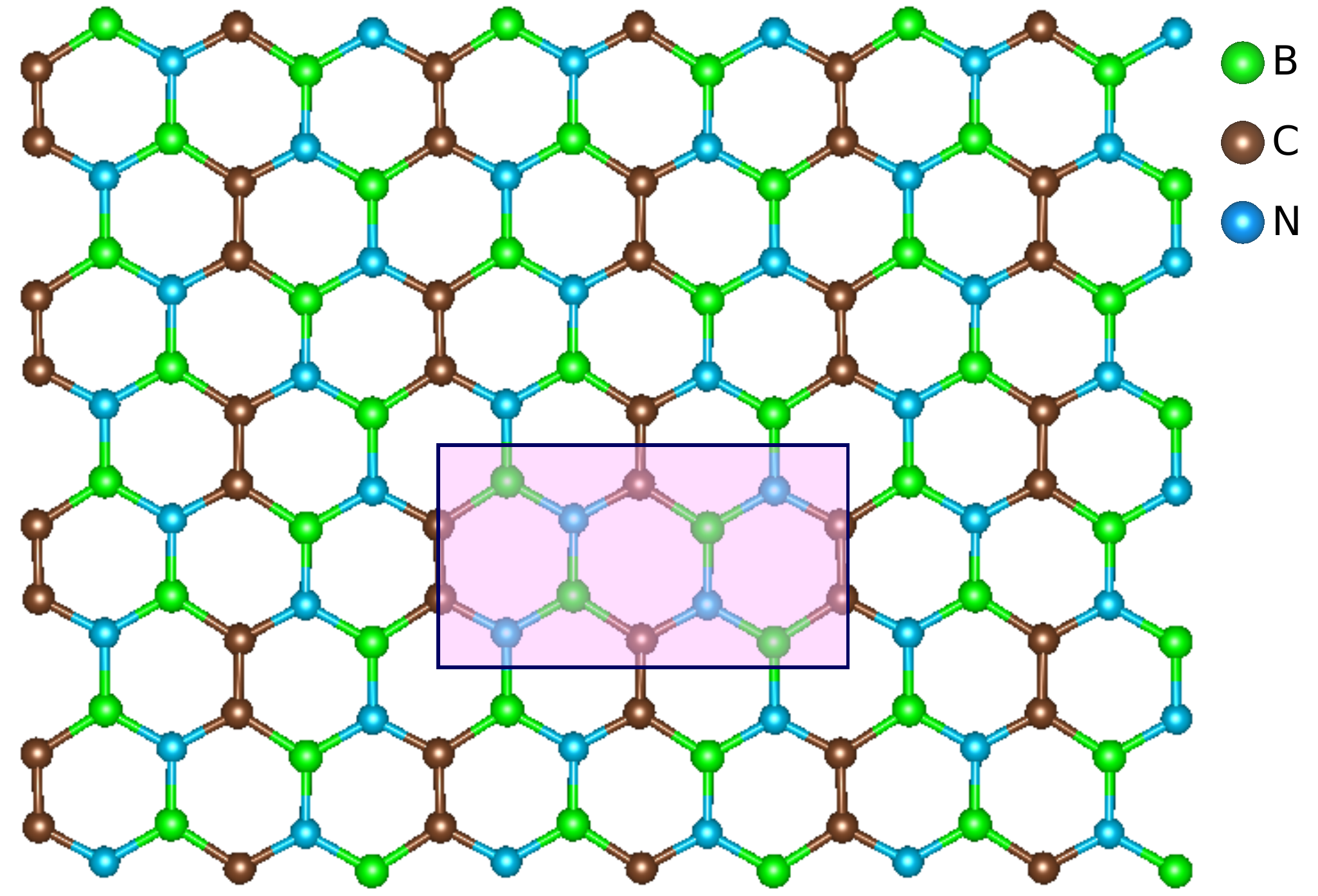}
    \caption{Schematic representation of the model ternary h-BCN structure studied here. Importantly, this model lattice refers to the most stable structure (named BCN\_v3) synthesized in the experiments reported in the reference \cite{beniwal2017graphene}. The area highlighted in this figure stands for the unity cell.}
    \label{fig:system}
\end{figure}

To study the electronic and structural properties of the ternary h-BCN lattice in the presence of lattice defects, we performed DFT calculations within the framework of generalized gradient approximation (GGA) with localized basis sets \cite{PhysRevLett.77.3865}, as implemented in the SIESTA code \cite{soler2002siesta}. The exchange-correlation functional used is based on the Perdew–Burke–Ernzerhof (PBE) scheme \cite{PhysRevLett.80.891}. To treat the electron core interaction, we used the Troullier–Martins norm-conserving pseudopotentials \cite{PhysRevB.64.235111}. We also include polarization effects and the Kohn–Sham orbitals are expanded with double-$\zeta$ basis \cite{PhysRevB.64.235111}. The structural relaxation of all model ternary h-BCN lattices studied here is carried out until the force on each atom is less than $10^{-3}$ eV/\r{A} and the total energy change between each self- consistency step achieves a value less or equal to $10^{-5}$ eV. The Brillouin zone is sampled by a fine $21\times 21\times 1$ grid and, to determine the self-consistent charge density we use a mesh cutoff of 200 Ry. A supercell geometry was adopted with a vacuum distance of 30 \r{A} to avoid interaction among each structure and its images. Figure \ref{fig:system} illustrates the model ternary h-BCN lattice studied here. Importantly, the model lattice presented in this figure refers to the most stable structure (named BCN\_v3) synthesized in the experiments reported in the reference \cite{beniwal2017graphene}. Our model h-BCN lattices contain 144 and 143 atoms for the non-defective and defective cases, respectively. 

\section{Results}

\begin{figure}[!htb]
    \centering
   \includegraphics[width=\linewidth]{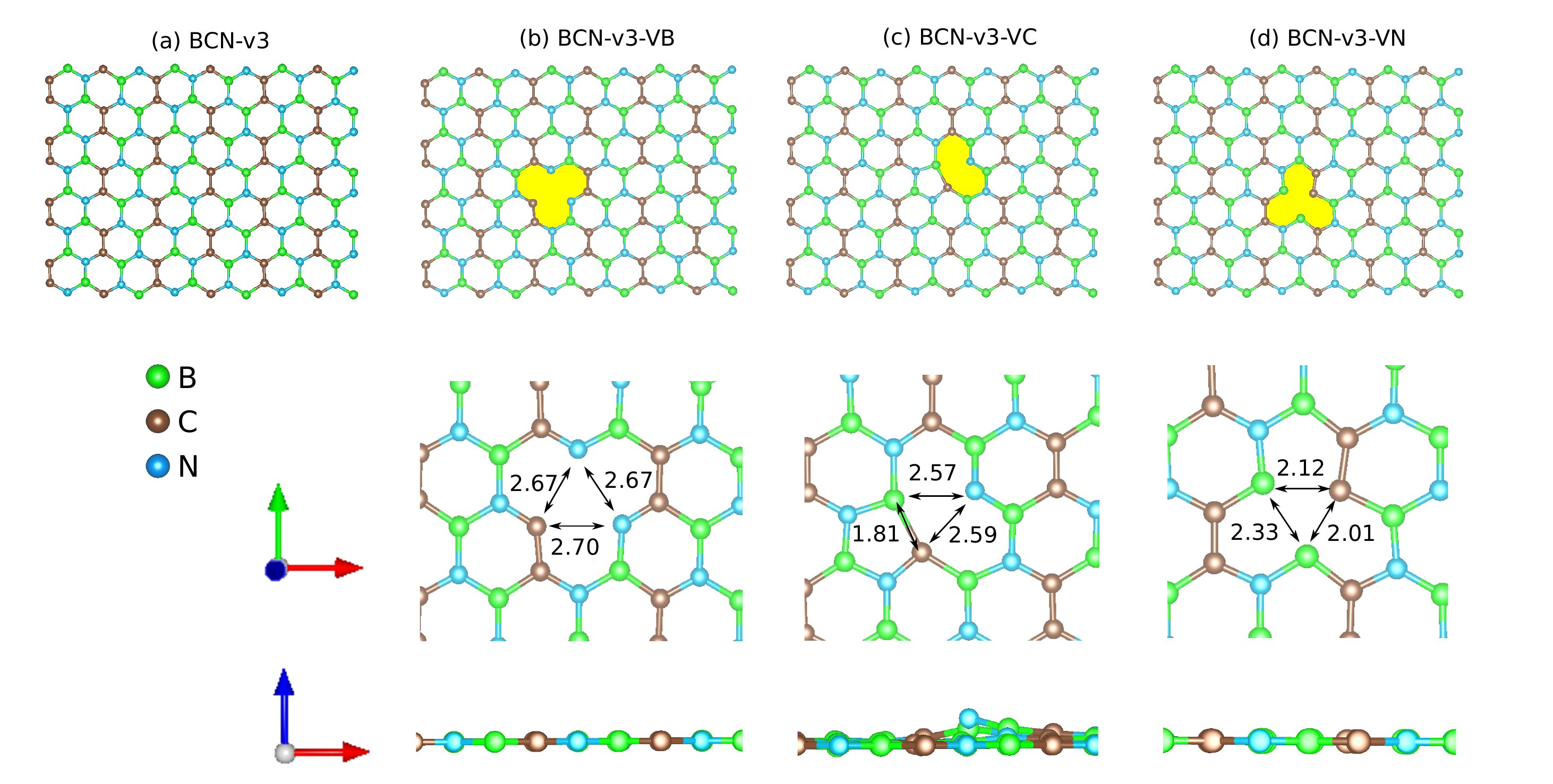}
   \caption{Ground state configurations for the (a) non-defective case (BCN-v3) and for the cases with (b) a single-atom vacancy of boron (BCN-v3-VB), (c) carbon (BCN-v3-VC), and  (d) nitrogen (BCN-v3-VN).}
   \label{fig:lattices}
\end{figure}

We begin our discussions by showing in Figure \ref{fig:lattices} the ground state configurations for the non-defective case (Figure \ref{fig:lattices}(a) -- BCN-v3) and for the cases with a single-atom vacancy of boron (Figure \ref{fig:lattices}(b) -- BCN-v3-VB), carbon (Figure \ref{fig:lattices}(c) -- BCN-v3-VC), and nitrogen (Figure \ref{fig:lattices}(d) -- BCN-v3-VN). The middle sequence of panels in Figure \ref{fig:lattices} displays enlarged images of the vacancy regions, and the bottom panels present side views of the lattices. In Figure \ref{fig:lattices}(a), the C--C, C--B, C--N, and B--N are 1.38 \r{A}, 1.54 \r{A}, 1.40 \r{A}, and 1.44 \r{A}, respectively. As can be inferred from the distances between the atoms in the vacancy regions, as depicted in the panels in the middle of Figure \ref{fig:lattices}, the BCN-v3-VB case (Figure \ref{fig:lattices}(b)) is the one with the smallest tendency of bond reconstructions. For this defective lattice, the distances between the atoms in the vacancy are almost the same. On the other hand, the BCN-v3-VC case (Figure \ref{fig:lattices}(c)) is the one with the higher tendency of bond reconstructions, in which a new C--B bond of 1.81 \r{A} was formed. This bond reconstruction yielded a new pentagonal ring, and a local deviation in the lattice planarity was also observed. The BCN-v3-VN case showed a considerable tendency for bond reconstructions, as denoted by the distances between the atoms in the vacancy, but no new bond was formed, as shown in Figure \ref{fig:lattices}(d). 

It is worthwhile to mention that in the BCN-v3-VC and BCN-v3-VN cases, the distances between the carbon and boron atoms in the center of the vacancy are smaller than the ones between carbon and nitrogen atoms when it comes to the BCN-v3-VB case. This trend is expected due to the higher number of electrons of nitrogen when contrasted with boron, which potentiates the repulsive effects in the vacancies with a higher number of the former. Moreover, the BCN-v3-VC case is the only one that presents three different atoms in the center of the vacancy, which considerably increases the degree of symmetry braking promoted by removing a carbon atom. This symmetry breaking allows the reconstruction of the B--C bond, as mentioned above. Note that in the BCN-v3-VB and BCN-v3-VN cases, two of the central atoms of the vacancies have the same chemical species, which contributed to keeping the symmetry of the bond lengths nearby the defect region.

The energetic stability of these structures was verified through cohesion energy calculations. To do so, the equation $E_{coh} = (E_{totoal} -N_bE_b-N_cE_c-N_nE_n)/N_{total}$ was employed, where $E_{total}$ is the total energy of the system and $N_v$ and $E_v$ are the number of atoms and total energy of an isolated atom $v$, respectively. In this case, $v$ stands for $B$, $C$, or $N$. The values cohesion energy obtained here were: -7.35 eV, -7.30 eV, -7.31 eV, and -7.32 eV for BCN-v3,  BCN-v3-VB,  BCN-v3-VC, and  BCN-v3-VB, respectively. One can note that these cohesion energy values are very close. It suggests that our model h-BCN defective lattices are energetically stable.

\begin{figure}[!htb]
    \centering
    \includegraphics[width=0.8\linewidth]{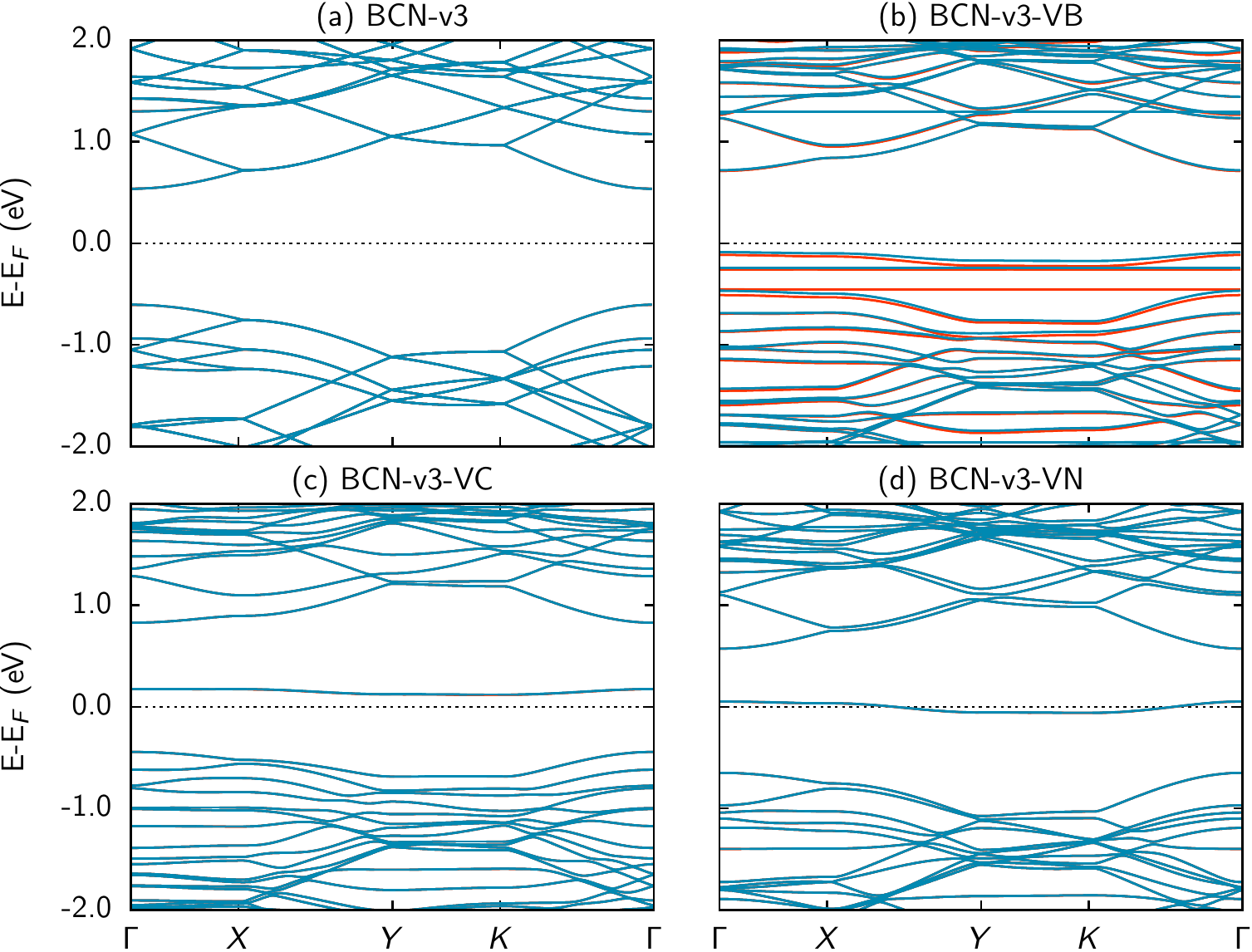}
    \caption{Calculated band structures for the (a) BCN-v3, (b) BCN-v3-VB, (c) BCN-v3-VC, and (d) BCN-v3-VN cases. The dashed horizontal line indicates the Fermi energy level.}
   \label{fig:bandgaps}
\end{figure}

We consider now the electronic properties of the model lattices studied here. Figures \ref{fig:bandgaps}(a), \ref{fig:bandgaps}(b), \ref{fig:bandgaps}(c), and \ref{fig:bandgaps}(d) illustrate the calculated band structure for the BCN-v3, BCN-v3-VB, BCN-v3-VC, and BCN-v3-VN cases, respectively. The bandgaps found for these cases were 1.14 eV, 1.17 eV, 1.17 eV, and 1.22 eV, respectively. Note that these values are in agreement with the experimental ones reported in the reference \cite{beniwal2017graphene}, which vary between 1.0-2.0 eV. In the band structure of all defective cases, one can note the appearance of flat midgap states as a consequence of the inclusion of vacancies. In Figures \ref{fig:bandgaps}(a), \ref{fig:bandgaps}(c), and \ref{fig:bandgaps}(d) one can see that the calculations yielded nonmagnetic ground state geometries. In the defective cases (Figures \ref{fig:bandgaps}(c) and \ref{fig:bandgaps}(d)), only one flat state appears within the bandgap. Figure \ref{fig:bandgaps}(b) shows that the band structure of the ternary h-BCN lattice is strongly affected in the presence of a single boron atom vacancy. In this case, some midgap states appear within the bandgap. One can realize a symmetry breaking of the density of spin-up (blue lines) and spin-down (red lines) electrons. This spontaneous magnetization yielded a magnetic dipole moment of about 0.87 $\mu_B$. The absence of a boron atom breaks the bond patterns of two nitrogen atoms, which generates a local spin moment. This trend was not observed in the other cases studied here.            

\begin{figure}[!htb]
    \centering
    \includegraphics[width=0.8\linewidth]{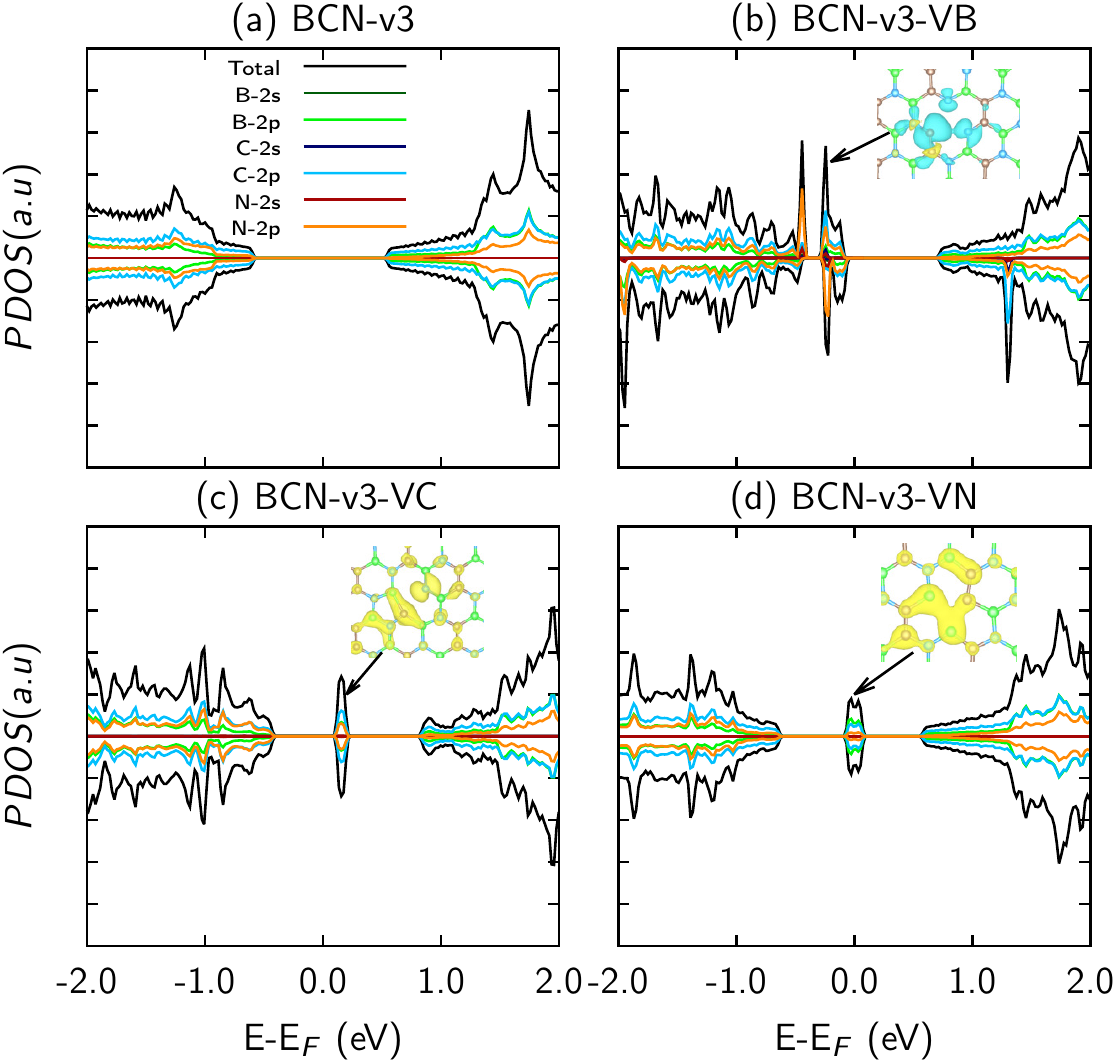}
    \caption{Corresponding projected density of states (PDOS) for the band structures presented in Figure \ref{fig:bandgaps}. The inset panels show the local density of states (LDOS), calculated for the states indicated with the arrows. In panel (b), the blue and yellow spots are denoting the up and down electrons, respectively. In panels (c) and (d), the yellow spots denote the total LDOS.}
    \label{fig:pdos}
\end{figure}

Finally, Figure \ref{fig:pdos} illustrates the corresponding projected density of states (PDOS) for the band structures presented in Figure \ref{fig:bandgaps}. The inset panels in Figures \ref{fig:pdos}(b-d), \ref{fig:pdos}(c), and \ref{fig:pdos}(d) show the local density of states (LDOS), calculated for the states indicated with the arrows. The blue and yellow spots are denoting the spin-up and spin-down electrons, respectively. In the inset panels of Figures \ref{fig:pdos}(c) and \ref{fig:pdos}(d), just the spin-down LDOS were shown once no differences between the two spin channels were observed for these cases. As a general trend, the major contribution to the DOS comes from the p-orbital of carbon atoms. Moreover, one can note that the excess of spin is localized close to the vacancies. The spontaneous magnetization produced by removing a boron atom is reflected by the splitting in the PDOS peaks for different energy levels, as shown in Figure \ref{fig:pdos}(b). In this monovacancy case, the greater contribution for the PDOS peaks within the bandgap comes from 2p-orbitals of nitrogen atoms, which characterizes the confinement of the electronic states observed in the band structure of BCN-v3-VB case (see Figure \ref{fig:bandgaps}(b)).

\section{Conclusion} 

In summary, the electronic and structural properties of a novel ternary h-BCN material in the presence of single-atom (boron, carbon, or nitrogen) vacancies were investigated in the framework of density functional theory calculations. Results showed that the BCN-v3-VB case has the smallest tendency of bond reconstructions. On the other hand, the BCN-v3-VC case is the one with a higher tendency of bond reconstructions. In this case, a new C--B bond of 1.81 \r{A} was formed, yielding a new pentagonal ring. The BCN-v3-VN case showed a considerable tendency for bond reconstructions, as denoted by the distances between the atoms in the vacancy, but no new bond was formed. The energetic stability of these structures was verified through cohesion energy calculations. We obtained very close values for the cohesion energies among all the model lattice studied here. It suggests that our model h-BCN defective lattices are energetically stable. The bandgaps found for the BCN-v3, BCN-v3-VB, BCN-v3-VC, and BCN-v3-VN cases were 1.14 eV, 1.17 eV, 1.17 eV, and 1.22 eV, respectively. These values are in agreement with the experimental ones reported in the reference \cite{beniwal2017graphene}, which vary between 1.0-2.0 eV. In the defective BCN-v3-VC and BCN-v3-VN cases, only one flat state appears within the bandgap. On the other hand, the band structure of the ternary h-BCN lattice is strongly affected in the presence of a single boron atom vacancy. In the BCN-v3-VB case, some midgap states appear within the bandgap. It was also noted a symmetry breaking of the density of spin-up and spin-down electrons. This spontaneous magnetization yielded a magnetic dipole moment of about 0.87 $\mu_B$. The absence of a boron atom breaks the bond patterns of two nitrogen atoms, which generates a local spin moment. This trend was not observed in the other cases studied here. 

\section{Acknowledgments}
The authors gratefully acknowledge the financial support from Brazilian Research Councils CNPq, CAPES, and FAPDF and CENAPAD-SP for providing the computational facilities. W.F.G. gratefully acknowledges the financial support from FAP-DF grant 0193.0000248/2019-32. L.A.R.J. gratefully acknowledges the financial support from CNPq grant 302236/2018-0. R.T.S.J. gratefully acknowledges, respectively, the financial support from CNPq grant 465741/2014-2, CAPES grants 88887.144009/2017-00, and FAP-DF grants 0193.001366/2016 and 0193.001365/2016. L.A.R.J. gratefully acknowledges the financial support from DPI/DIRPE/UnB (Edital DPI/DPG $03/2020$) grant $23106.057541/2020-89$ and from IFD/UnB (Edital $01/2020$) grant $23106.090790/2020-86$.

\bibliographystyle{iopart-num}
\bibliography{references}

\end{document}